\documentclass[%
reprint,  
superscriptaddress,
nofootinbib,
amsmath,amssymb,
aps,
pra,
raggedbottom,
]{revtex4-1}

\usepackage{graphicx}
\usepackage{dcolumn}
\usepackage{bm}
\usepackage{lipsum}
\usepackage{float}
\usepackage{subfigure}

\usepackage[colorlinks=true,bookmarks=false]{hyperref}
\hypersetup{linkcolor=blue,citecolor=red,urlcolor=blue}   

\begin{document}

\preprint{APS/123-QED}

\title{Field Driven Charging Dynamics of a Fluidized Granular Bed}

\author{R. Yoshimatsu}
\affiliation{%
Computational Physics for Engineering Materials, IfB, ETH Zurich, Wolfgang-Pauli-Strasse 27, 8093 Zurich, Switzerland
}
\author{N.A.M. Ara\'ujo} 
\affiliation{%
Departamento de F\'isica, Faculdade de Ci\'encias, Universidade de Lisboa, P-1749-016 Lisboa, Portugal, and Centro de F\'isica Te\'orica e Computacional, Universidade de Lisboa, P-1749-016 Lisboa, Portugal
}
\author{T. Shinbrot} 
\affiliation{%
Department of Biomedical Engineering, Rutgers University, Piscataway, New Jersey, 08854, USA 
}
\author{H.J. Herrmann}
\affiliation{%
Computational Physics for Engineering Materials, IfB, ETH Zurich, Wolfgang-Pauli-Street 27, 8093 Zurich, Switzerland
}
\affiliation{%
Departamento de F\'isica, Universidade Federal do Cear\'a, 60451-970 Fortaleza, Cear\'a, Brazil
}
                  
\date{Febuary 2, 2015}
\begin{abstract}
A simplified model has previously described the inductive charging of colliding identical grains in the presence of an external electric field. Here we extend that model by including heterogeneous surface charge distributions, grain rotations and electrostatic interactions between grains. We find from this more realistic model that strong heterogeneities in charging can occur in agitated granular beds, and we predict that shielding due to these heterogeneities can dramatically alter the charging rate in such beds.
\end{abstract}
 
\maketitle

\section{\label{introduction}Introduction}
Granular materials have long been known to spontaneously develop strong charges, for example in volcanic plumes and in sandstorms\cite{baddeley, Anderson28051965, arason2011observations, rakov2003lightning, lacks2011contact, merrison2012sand,thomas2007electrical, chakraborty2009volcanic}. Granular charging is also important in many industries, such as in printing and pharmaceutical formulation\cite{cademartiri2012simple, verakis1987brief, mehrotra2007spontaneous, thyagu2012stuck}. Despite the importance and prevalence of granular charging, its underlying causes remain controversial.

Past studies have largely focused on geometric or material differences between grains\cite{lowell1986triboelectrification, terris1989contact, forward2009charge, kok2009electrification,  waitukaitis2014size}. Remarkably, however, grains that are absolutely identical in shape, size and chemical composition have also been found to charge one another\cite{shinbrot2008spontaneous,apodaca2010contact,shaw1929tribo,shaw1926electrical,miura2006gas,hu2012contact,kok2009electrification,lacks2008nonequilibrium}.

To explain the charging of identical grains, P{\"a}htz $et$ $al$. proposed a simplified model\cite{pahtz2010particle} in which an external electric field\cite{PhysRevX.5.011002} - as might be produced by a nearby electrical storm\cite{sow2011electrification} - can induce polarizations in grains. Zhang \textit{et al.} showed that when grains collide in the presence of an external field, the amount of exchanged charges does increase with the field strength\cite{Zhang15}. Experimental results by Lee \textit{et al.} clearly show that polarization plays an important role in the collective dynamics of grains, even in the absence of an external field\cite{Lee15}.

That model was simplified by considering only vertical dipole moments and by neglecting Coulomb forces between grains. Here, we refine the earlier model by including higher order electrical moments on grains and by allowing grains to interact through Coulomb forces. Additionally, the prior model assumed that the external field was overwhelmingly stronger than that due to nearby grain charges. We remove that simplifying assumption here as well. 

\begin{figure}[t]
\centering
\includegraphics[width=8.5cm]{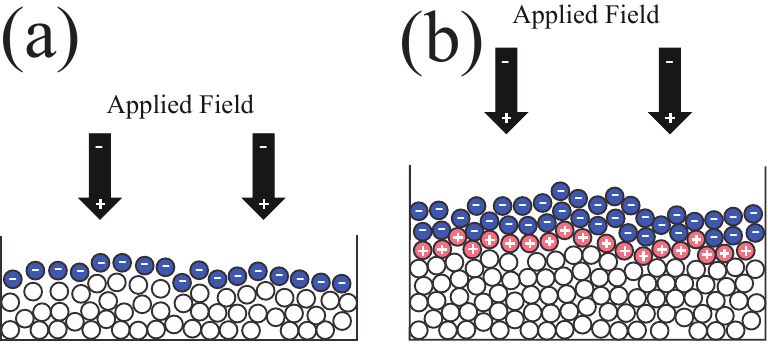}
\caption{Caricature of the expected charge heterogeneities. (a) For a shallow granular bed, we expect a monolayer of negative charges at the top of the bed and (b) for a thicker bed, we expect a charge inversion due to the field induced by the negative layer.}
\label{prediction}
\end{figure}

Our goals in performing these more detailed simulations are twofold. Prior work Ref. \cite{pahtz2010particle} showed a mechanism for pumping negative charges up and positive charges down in the presence of a downward external electric field, but neglected interactions between charged grains. So first we seek to include these interactions and evaluate how they affect subsequent grain charging. Second, the charging mechanism described in Ref. \cite{pahtz2010particle} should produce significant charge heterogeneity that has not been quantitatively examined. In particular, if the bottom of the bed is grounded, then positive charges will be drained from beneath the bed, leaving growing negative charges near the top of the bed, as sketched in Fig.\,\ref{prediction}(a). Significantly, we can expect these negative charges to shield the applied field within the bed and so to reduce granular charging. Moreover, as this shielding layer of negative charges grows sufficiently large, it can exceed the influence of the external field and induce its own charges within the bed - in this case of opposite, positive sign. This would produce a double layer, with negative charge at the top of the bed, overlying an induced positive charge, sketched in Fig.\,\ref{prediction}(b).

We therefore present a detailed simulation of granular charging both to provide a more accurate investigation into the mechanism of charge transfer than previously possible, and to quantify expected charge heterogeneities in granular beds. In section \ref{model} we describe this simulation, then in section \ref{results}, we analyze results of the simulation, and in section \ref{conclusion}, we draw conclusions.

\section{\label{model}Model}
\begin{figure}[t]
\centering
\includegraphics[width=8cm]{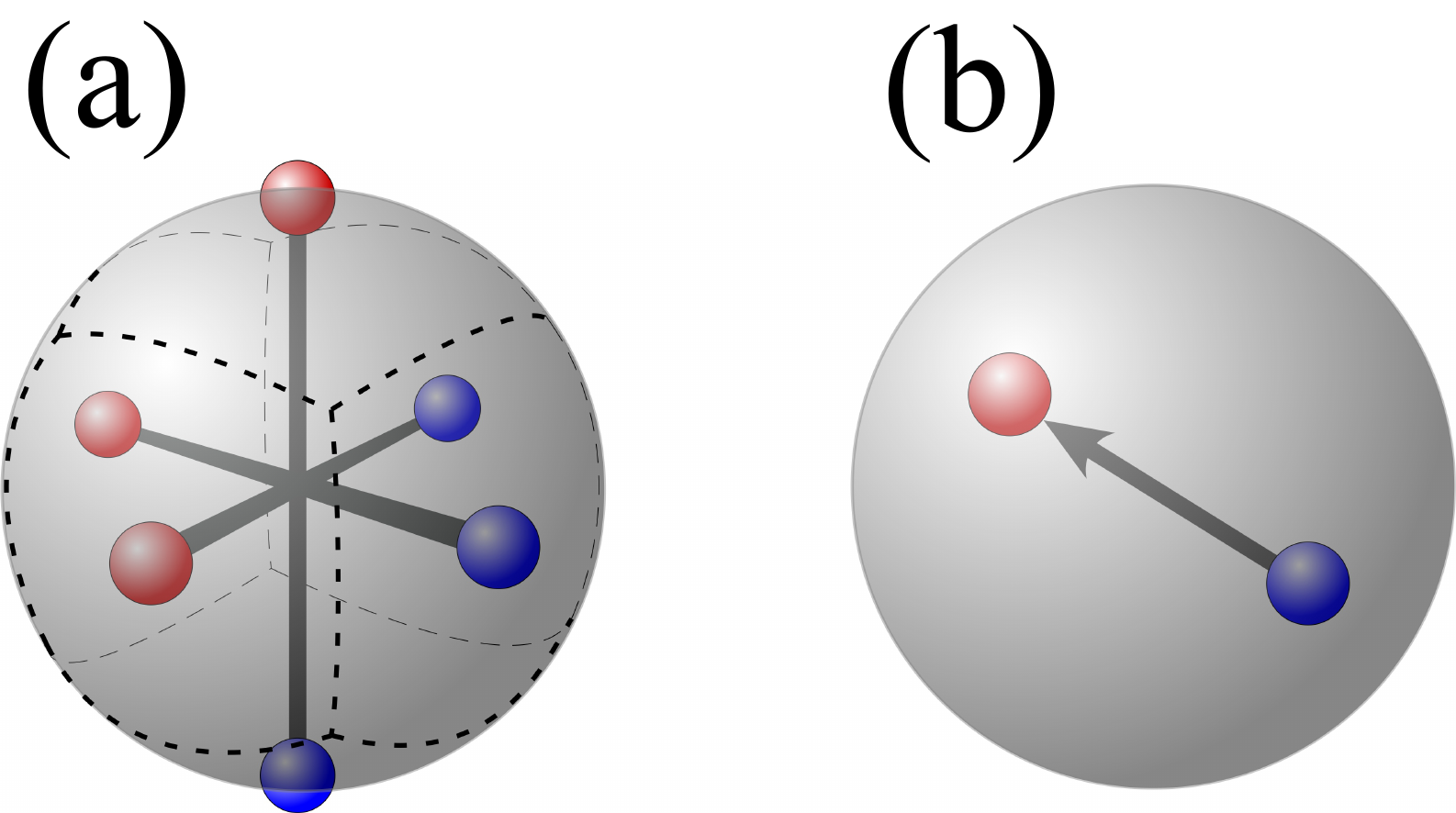}
\caption{(a) Schematic representation of a grain with six constituent charges.  Each charge is centered on a surface domain delineated here by broken lines. (b) Illustrative dipole moment calculated from charges on panel (a).}
\label{particle}
\end{figure}

Here, we provide a general description of the model. Details about the model and numerical simulations can be found in the Appendix. We simulate charging of grains in an agitated bed of spherical grains. Grains can accumulate charges on their surface. Since they are insulators, these charges are heterogeneously distributed. To account for such heterogeneity, we model the surface as six independent, orthogonally placed charges as sketched in Fig.\,\ref{particle}(a). This allows us to define complex electrical moments.

To prevent crystallization, we consider a polydisperse distribution of grain radii, $R_i$, following a Gaussian distribution with variance 10\% of the mean $\bar{R}$=0.75 mm. However, as we are interested in the dynamics of identical grains, we assume that all grains have the same mass, $m=4.239\times 10^{-6}$ kg, corresponding to a glass of average radius  $\bar{R}$ and density $\rho_{g}=2.4\times10^{3}$ kg/m$^{3}$.
 
\underline{Equations of motion:} Numerically, we solve the equations of motion for each grain by means of the discrete element method described elsewhere \cite{granular}. We track both translational and rotational motions of the $i^{th}$ grain and include all mechanical and electrostatic forces and torques:
\begin{align}
\begin{split}
	m\frac{d\vec{v}_i}{dt}=&[\vec{F}_{g}+ \vec{F}_{ela,i}+\vec{F}_{f,i}]+ \vec{F}_{ele,i},\\
	I_i\frac{d\vec{\omega}_i}{dt}=&[\vec{T}_{m,i}]+\vec{T}_{e,i}.
\end{split}
\label{equation_of_motion}
\end{align}

\underline{Mechanical forces:} The mechanical terms, in square brackets, are as follows. The gravitational force, $\vec{F}_g$, is defined in the usual way using earth's gravity. For the elastic force, $\vec{F}_{ela,i}$, we use the model of Walton and Braun \cite{walton1986viscosity} with restitution coefficient 0.935. This value is the same as was used by Ref. \cite{pahtz2010particle}, and is deliberately chosen to be large because we agitate beds of varying depths from below (described shortly). Lower restitution coefficients require either that we increase the agitation strength with bed depth or that we maintain strong constant agitation. The first alternative introduces a new parameter that changes with depth, while the second produces very different states for shallow and deep beds. Neither is desirable, whereas by choosing a high restitution coefficient, we are able to produce a nearly uniformly colliding state for all beds studied without changing the agitation strength \cite{gallas1992molecular}. We use a standard kinetic friction model for $\vec{F}_{f,i}$ \cite{halliday2010fundamentals}. As for the torque equation, $I_i$ is the moment of inertia of the $i^{th}$ grain, and the mechanical torque, $\vec{T}_{m,i}$, is determined from the kinetic friction at the contact point of two colliding grains \cite{halliday2010fundamentals}.

We agitate the bed by using a "splash function" to re-inject grains that hit the bottom of the computational domain. Such grains acquire velocity $\vec{V}^{s}$=2.7$\sqrt{2g\bar{R}}\hat{z}$, which has long been used to model saltation and fluidization \cite{anderson1988simulation}.

\underline{Electrostatic forces:} The electrostatic terms in Eq.\,\eqref{equation_of_motion} are calculated as follows. We track all six charges shown in Fig.\,\ref{particle}(a) for every grain, however, instead of a fully detailed calculation of the electrostatic interactions\cite{Nakajima99}, we only use mono and dipole terms to compute Coulomb forces, $\vec{F}_{ele,i}$, and electrostatic torques, $\vec{T}_{e,i}$. The mono and the dipole terms are defined in the obvious way, so that the net charge is the sum of all six constituent charges and the dipole moment of the $i^{th}$ grain is:
	\begin{align}
	\vec{d}_i & =  \sum_{n=1}^{6} q_{i,n}\vec{r}_{i,n},
	\label{dipole}
	\end{align} 
where $q_{i,n}$ is the $n^{th}$ charge of the $i^{th}$ grain and $\vec{r}_{i,n}$ is the position vector of this charge measured from the grain center. We avoid far-field dipole approximations and calculate all Coulomb forces exactly by evaluating $q_{1}q_{2}/r_{12}^2$ for all charges, $q_{1}$ and $q_{2}$, and all distances $r_{12}$ between these charges. We prevent divergence when two grains come into contact by defining dipole forces and torques via two virtual charges displaced by $\frac{2}{3}R_i$ from the center of the grain, as sketched in Fig.\,\ref{particle}(b). Each virtual charge $q^d_i$ is obtained from:
\begin{align}
	q^d_i =\frac{3 |\vec{d}_i|}{4 R_i}.
\label{dipole_charge}
\end{align} 

We account for long-range electrostatic interactions between charged grains by using the Particle-Particle Particle-Mesh (PPPM) method described elsewhere \cite{hockney1988computer}. Details are included in the appendix, but in short, through this method we subdivide the computational volume into a cubic grid, where the linear dimension of each grid cell is 4$\bar{R}$, and we assign the net charge of each grain to its closest grid intersection. We duplicate the computational domain ten times in both horizontal directions (for a total of 440 duplicates surrounding a central domain), and we use the electric field, $\vec{E}_i$, from these duplicates to calculate Coulomb forces and induced charges (described next) on each grain. In this way, we produce nearly periodic boundary conditions in a finite computational domain.

\underline{Charge transfer:} We simulate charge transfer similarly to Ref. \cite{pahtz2010particle}: each grain receives an induced polarization proportional to the electric field that it is subjected to, and charges can be transferred between grains due to neutralization events during contact. In detail, a local electric field of amplitude $E_{i}$ at the center of the $i^{th}$ grain, calculated by the PPPM method, induces a dipole moment: 
\begin{align}
\vec{P}_{i}=\alpha \vec{E}_{i}.
\label{polarization}
\end{align}
Here $\alpha$ is the polarizability in cgs units. Unlike Ref. \cite{pahtz2010particle}, $\vec{E}_i$, is not only the applied external electric field but also includes the field computed using the PPPM method due to surrounding grains. The induced dipole moment is assigned to the nearest of the three pairs of charges shown in Fig. \,\ref{particle}(a). In order to compute neutralization in a well-defined and charge-conserving manner, during collision permanent charges on the contacting sectors neutralize according to:
\begin{align}
q_{i}^{after}=q_{j}^{after}= \frac{1}{2}[q_{i}^{before}+q_{j}^{before}],
\label{neutralization}
\end{align}
where $q_{i}$ and $q_{j}$ are the charges on contacting domains of the $i^{th}$ and $j^{th}$ contacting grains. During a collision, we add the induced dipole charges to the nearest sector of the three pairs of charges to produce new permanent charges.

Finally, we mention three technical points needed to close the description of the simulation. First, we model granular charging using a grounded bottom surface. This is the boundary condition used in Ref. \cite{pahtz2010particle}, which permits the injection of charge into the bed to mimic field and laboratory observations \cite{sow2011electrification,zheng2006effect,zhou2002experimental,zheng2003laboratory}. Second, to prevent spurious repetition of charging, we only apply induction and neutralization operations (Eqs.\,\eqref{polarization} and \eqref{neutralization}) once per contacting grain pair, at the moment when the grains separate. Third, the algorithm that we have described applies to binary collisions. In rare cases when a grain simultaneously loses contact with multiple neighbors, we perform neutralization sequentially in random order. 

\section{\label{results}Results}
\begin{figure}[t]
\centering
\includegraphics[width=8.5cm]{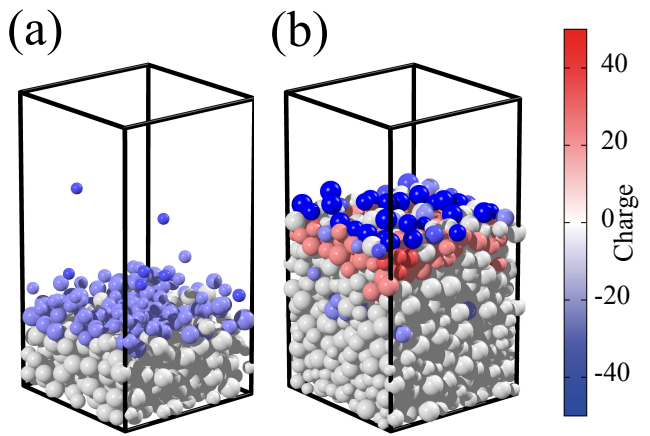}
\caption{Visualization of the bed with (a) nominal bed depth $n_{L}=5$ and (b) $n_{L}=13$ in asymptotic states. Grains are colored according to their net charge, where a unit charge corresponds to $1.3\times10^{-12}$ C. One particle layer here consists of approximately 100 grains.}
\label{electrostatics}
\end{figure}

Using this simulation that we have described, we first evaluate the extent to which heterogeneities in the bed appear as discussed in the introduction, and we second dissect the mechanism of granular charging in greater detail than previously possible.

\underline{Charge heterogeneities:} Qualitative assessment of expected charge heterogeneities can be seen from Fig. \ref{electrostatics} , where we visualize the granular beds in their asymptotic states. For a shallow bed, we see negative charges accumulating at the top, and for a deeper bed, we see strong negative charges at the top, overlying positive charges.

\begin{figure}[t]
\centering
\includegraphics[width=8cm]{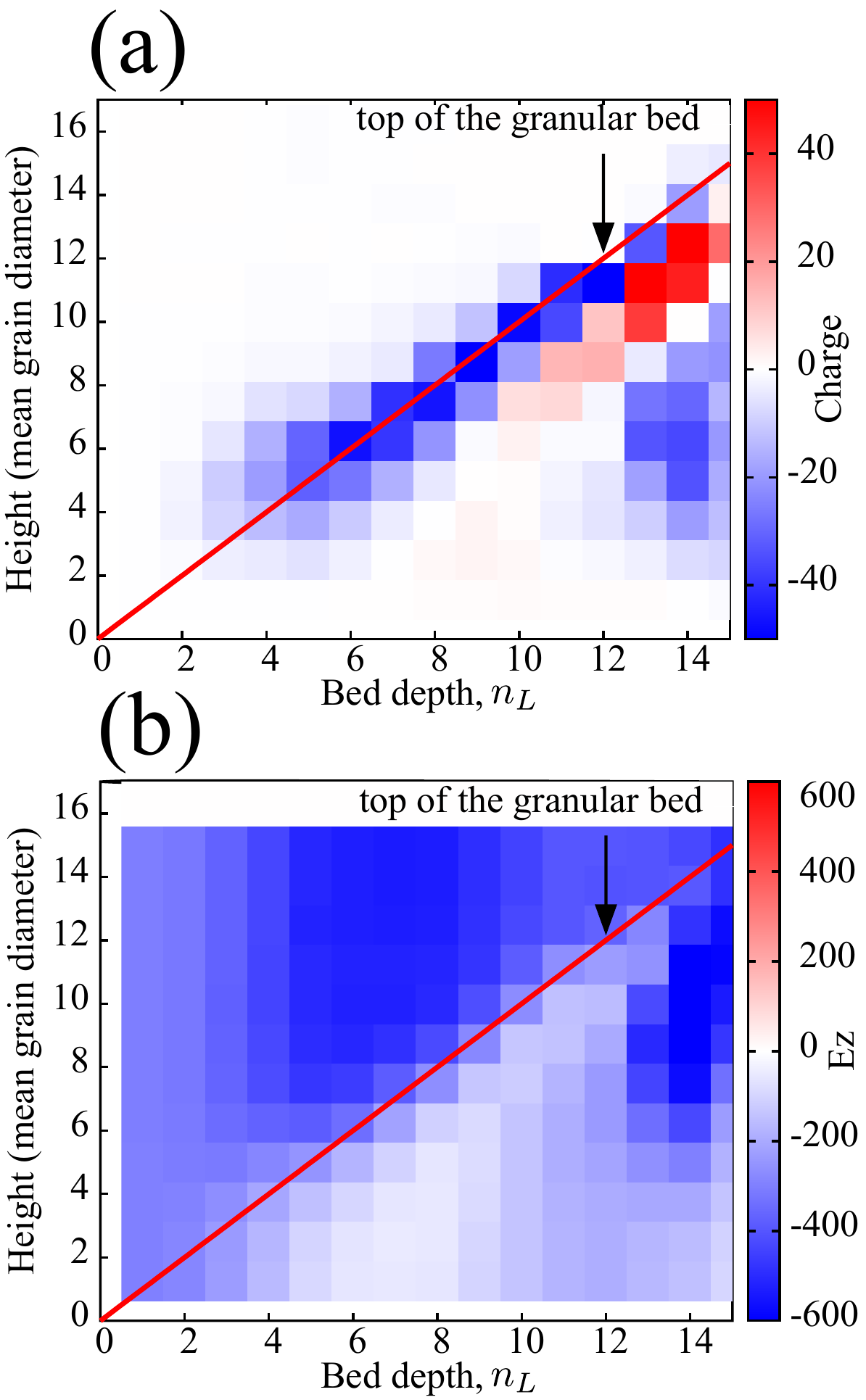}
\caption{(a) Net charge in different height bins (see text) versus bed depth. Red line shows top of granular bed and cells are colored according to mean charge of grains within each bin. Color bar shows charge magnitude, 1 unit charge is 1.3$\times$$10^{-12}$ C. (b) Vertical component of field versus bed depth. The applied vertical external field is 300 kV/m pointing down, and color bar shows vertical field magnitude in kV/m. Red line again indicates top of granular bed.}
\label{charge_field}
\end{figure}

We quantitatively assess these observations by performing independent simulations for different bed depths. We then bin charge and electric field values as a function of height, as shown in Fig.\,\ref{charge_field} where we display charge and field values as a function of bed depth, $n_{L}$. To bin charges and fields, we divide the computational domain into horizontal slices of thickness $2\bar{R}$, and integrate the charge or the vertical component of the field over each slice. In agreement with  Fig.\,\ref{electrostatics}, Fig.\,\ref{charge_field}(a) shows that for $n_{L}\geq3$, a layer of negatively charged grains forms at the top of the bed, and Fig.\,\ref{charge_field}(b) confirms that the field beneath this layer drops significantly. The horizontal components of electric field are invariably two orders of magnitude smaller than the vertical component, $Ez$, so we report only $Ez$ in Fig.\,\ref{charge_field}(b). Once the bed depth exceeds about 9 grain diameters, a second layer of positively charged grains emerges, shown in red in Fig.\ref{charge_field}(a). There is also a suggestion of a third layer, of negative charges, beneath this. These results suggest that, as the number of layers increases, the top layer of negative charges eventually exceeds the influence of the external field and it induces charges of opposite sign beneath it, as shown in Fig.\,\ref{electrostatics}.

Thus our results appear to confirm expectations that there should be significant electrical heterogeneity in agitated granular beds, with charges concentrating near the top of the bed, and electric fields within the bed being strongly shielded by these charge concentrations. More than this, as we have mentioned our simulations are more detailed than those performed previously, and so we can probe the essential mechanism that Ref. \cite{pahtz2010particle} sought to elucidate, namely how charging of identical grains occurs and what it depends on. Here we find a surprise.

\begin{figure}[t]
\centering
\includegraphics[width=7.5cm]{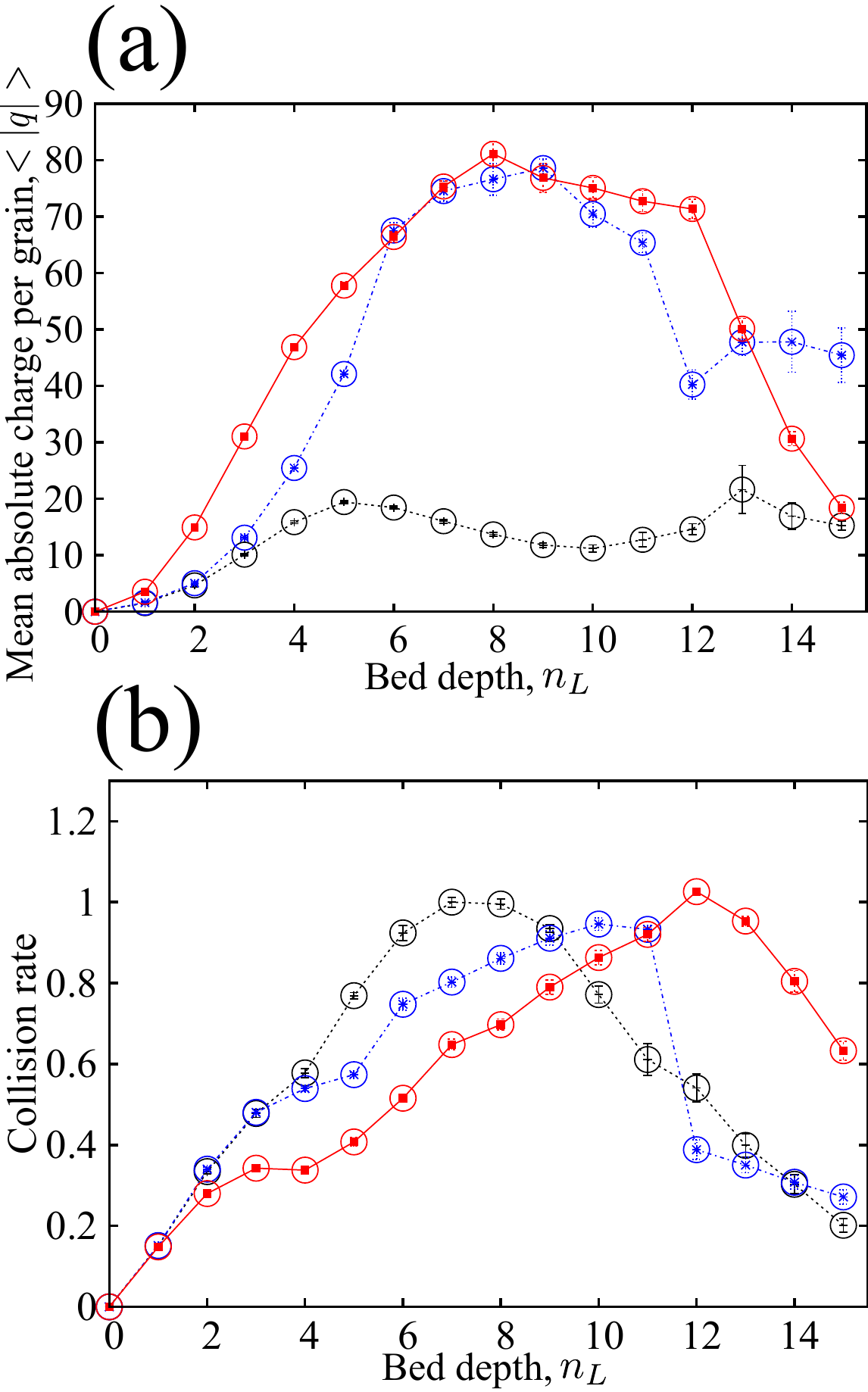}
\caption{(a) $<|q|>$ vs. bed depth in several scenarios. Red: simulations lacking multipoles, Coulomb forces, and field-dependent polarization (similar to Ref. \cite{pahtz2010particle} ); blue: simulations with multipoles that rotate with grains, but lacking Coulomb forces and field-dependent polarization; black: full simulations described in text. (b) Rate of grain collision events vs. bed depth in each of these scenarios. Each data point is an average of 200 measurements taken at sequential times in the asymptotic state and the error bars show the standard errors of the mean (smaller than the symbols).}
\label{mcg}
\end{figure}

\underline{Charging mechanism:} We investigate the charging mechanism by performing our simulations under strategically differing conditions. First, we repeat the results of Ref. \cite{pahtz2010particle} at multiple bed depths by excluding multipoles on grains, Coulomb forces between grains, and field-dependent polarization (Eq.\,\eqref{polarization}). Second, we include multipoles and repeat the same simulations, and finally we run the full simulation that we have described up to this point. In all cases, we evaluate the grain charging, explicitly the mean absolute value of the grain charge, $<|q|>$. Results are shown in Fig.\,\ref{mcg}.

Red symbols in Fig.\,\ref{mcg} show our simulation for conditions of Ref. \cite{pahtz2010particle}, blue symbols include multipoles that rotate with the grains, and black symbols indicate the full simulations. 

From Fig.\,\ref{mcg}(a), we see that including multipoles has little effect: simulations with dipoles oriented only in $\hat{z}$ direction (red), and those with multipoles that rotate with grains (blue) produce largely similar charging behaviors.  This is unexpected: one might anticipate that rotating dipoles would orient in arbitrary directions, as likely building as diminishing ultimate charging, but in fact this only slightly reduces ultimate charges on grains. From this perspective, prior work using a simplified model appears to be essentially unchanged by more careful simulations. On the other hand, including Coulomb forces and field-dependent charge induction (black) dramatically reduces charging rates, by as much as a factor of five. This again is unanticipated.

To establish the cause of this reduction, we recall that it has previously been determined (Ref. \cite{pahtz2010particle}) that charging is proportional to collision rate. It is therefore plausible that Coulomb forces could prevent collisions between like-charged particles and so could lead to the reduced charging shown in Fig.\,\ref{mcg}(a).  To assess this possibility, we evaluate the collision rate, summed over each of our simulations, and plot these in Fig.\,\ref{mcg}(b). Evidently the collision rate reaches a maximum at different bed depths for each simulation scenario, but each scenario reaches the same maximum, and at $n_L$=8-10, where from Fig.\,\ref{mcg}(a) we see that granular charging differs most dramatically between the full simulation and its simplified cousins, collision rates shown in Fig.\,\ref{mcg}(b) are nearly identical for all three cases. Evidently then, changes in collision rates due to Coulomb forces cannot account for the change in charging seen in Fig.\,\ref{mcg}(a).

Thus including Coulomb forces has a weak effect on collision rate, but this does not account for the reduction in bed charging seen. On the other hand, we have already seen from Fig.\,\ref{charge_field}(b) that the bed is strongly shielded from applied fields by the overlying charge layers shown in Fig.\,\ref{charge_field}(b).  Since the more complete model represented by black symbols in Fig.\,\ref{mcg}(a) generates charging \textit{in proportion to local electric fields}, it seems likely that this shielding effect is responsible for reduced charging.

\section{\label{conclusion} Conclusion}
We have performed detailed simulations of charging in agitated granular beds. These simulations confirm that in the presence of a vertical external electric field, colliding grains pump charge from a grounded bottom surface to the top of the bed. The simulations also reveal that the charges at the top of the bed can grow until they shield the interior of the bed from the external field. At this point, logically enough, granular charging is suppressed. We find that although the essential mechanism previously investigated continues to function in the presence of the more detailed considerations that we have described, the shielding observed can reduce ultimate granular charge levels by as much as a factor of five from those calculated using a more simplified model. We have also found that significant charge heterogeneities can establish themselves in agitated beds that are exposed to external fields, and we anticipate that future studies into heterogeneous charge distributions may provide insights into both charging and discharging dynamics in natural and industrial granular flows.

\section{\label{acknowledgment} Acknowledgment}
We acknowledge financial support from the ETH Grant, the ETH Risk Center, the Brazilian institute INCT-SC, and grant number FP7-319968 of the European Research Council. NA acknowledges financial support from the Portuguese Foundation for Science and Technology (FCT) under Contracts nos.  EXCL/FIS-NAN/0083/2012, UID/FIS/00618/2013, and IF/00255/2013. TS acknowledges support from the NSF DMR, award $\sharp$1404792.

\bibliography{paper}
\clearpage

\appendix*
\section{\label{appendix}Model details}
We present details of the simulations used here.

\underline{Time step:} The time step for the numerical integration is set to 5$\times$10$^{-5}$ seconds. This produces, for the most energetic grains, a minimum of more than 100 time steps per collision, allowing the dynamics to be stable.

\underline{Elastic force:} The elastic force acting on the $i^{th}$ grain resulting from a collision with the $j^{th}$ grain is given by:
\begin{equation}
    \centering
       \vec{F}_{ela,ij} =
        \begin{cases}
            0, & \vec{\epsilon}_{ij}=0 \\
            k_{l} \vec{\epsilon}_{ij}, & \frac{d |\vec{\epsilon}_{ij}|}{dt} \geq 0\\
            k_{u} \vec{\epsilon}_{ij}, & \frac{d |\vec{\epsilon}_{ij}|}{dt} < 0\\
        \end{cases}
    \label{walton_braun}
\end{equation}
where $\vec{\epsilon}_{ij}$ is the overlapping vector defined as:
\begin{equation}
    \centering
      \vec{\epsilon}_{ij} =
        \begin{cases}
            0, & \text{if } (R_{i}+R_{j}-|\vec{r}_{ij}|) \leq 0\\
           (R_{i}+R_{j}-|\vec{r}_{ij}|) \hat{r}_{ij}, & \text{otherwise}.
        \end{cases}
    \label{overlap}
\end{equation}
Here, $\vec{r}_{ij}$ is the vector connecting the centers of the two grains from $j$ to $i$. $k_l$ and $k_u$ are the elastic coefficients when the colliding grains are approaching or moving away from one another, respectively. We use $k_l=0.07$ and $k_u=0.08$, thus fixing the restitution coefficient to $\sqrt{k_{l}/k_{u}}=\sqrt{0.07/0.08}\approx0.935$, as described in the text.

\underline{Frictional force:} The frictional force acting on the $i^{th}$ grain resulting from a collision with the $j^{th}$ grain is given by:
\begin{align}
\vec{F}_{f,ij}=-\mu_{k} |\vec{F}_{ela,ij}| \hat{v}_{ij},
\end{align}
where $\mu_{k}=0.4$ is the kinetic friction coefficient, and $\hat{v}_{ij}$ is the unit vector with the direction of the relative velocity of the contact points of two colliding grains:
\begin{align}
\vec{v}_{ij}=(\vec{v}_i + \vec{\omega}_i \times \vec{R}_{cont,i}) -(\vec{v}_j + \vec{\omega}_j \times \vec{R}_{cont,j}),
\end{align}
where $\vec{R}_{cont,i,j}$ are the position vectors of the contact point measured from center of the $i^{th}$ and the $j^{th}$ grains.

\underline{Particle-Particle Particle-Mesh (PPPM) method:} The PPPM method is used to calculate the electrostatic interactions. As described in the text, the computational domain is divided into 3D grid cells, and net charges of the grains are assigned to the closest intersections between cells. Accordingly, the total charge assigned to the $k^{th}$ grid intersection is:
\begin{align}
Q_{k}=\sum_{i=1}^{N} q_{i} W(\vec{r}_{k}-\vec{r}_{i}),
\end{align}
where $N$ is the total number of grains, $q_i$ is the net charge of the $i^{th}$ grain, $\vec{r}_{k}$ is the position vector of the $k^{th}$ grid intersection and $\vec{r}_{i}$ is the position vector of the center of the $i^{th}$ grain. $W(\vec{r}_{k}-\vec{r}_{i})$ is the charge assignment function defined as:
\begin{equation}
       W(\vec{r}_{k}-\vec{r}_{i}) =
        \begin{cases}
   	 1, \text{if }& -2\bar{R} \leq (x_{k}-x_{i}) < 2\bar{R},\\
	 & -2 \bar{R} \leq (y_{k}-y_{i}) < 2\bar{R}\\
	  & \text{and }-2\bar{R} \leq (z_{k}-z_{i}) < 2\bar{R}\\
          0, & \text{otherwise}\\
           \end{cases}
    \label{charge_assignment}
\end{equation}
where $4\bar{R}$ is the linear length of the cubic cell. The electric field at the $k^{th}$ grid intersection is:
\begin{equation}
\begin{split}
	\vec{E}_{k}=k_e  \sum_{l \neq k}^{M} Q_{l}\frac{\vec{r}_{kl}}{|r_{kl}|^3},
\end{split}
\end{equation}
where $k_e$ is the Coulomb constant, $\vec{r}_{kl}$ is the vector connecting the two grid intersections from $l$ to $k$. For long-range interactions, we approximate the field at the center of a grain to be the field at the closest grid intersection. We include torques on dipoles due to electrostatic interactions beyond a grain's grid cell and its nearest neighboring cells, and as described in the text, we include all interactions within this domain, however we neglect Coulomb forces due to repeated charges outside of this domain because the dipole moment decays rapidly, with $\frac{1}{r^3}$. Coulomb forces for grains within any grid and its nearest neighbors are calculated in the usual way, according to:
\begin{equation}
\vec{F}_{ele,ij}=k_{e} q_{i,1} q_{j,2} \frac{\vec{r}_{12}}{|\vec{r}_{12}|^3}
\label{eq.electro}
\end{equation}
where $q_{i,1}$ and $q_{j,2}$ are all net or dipole charges of both grains and $\vec{r}_{12}$ is position vectors between these charges. Coulomb forces including both short- and long- range terms are calculated according to:
\begin{equation}
\vec{F}_{ele,i}=q_{i} \vec{E}_{ex}+ \sum_{i \neq j}^{n.n.}\vec{F}_{ele,ij}+q_{i} \vec{E'}_{k},
\label{eq.electrostatic}
\end{equation}
where the first term is the net charge coupling with the applied external field, $\vec{E}_{ex}$, the second term is the short-distance interaction with the grains within the same or nearest neighbor grid cells and the last term is the long-range interaction with distant grid intersections. To prevent double-counting, we subtract the contribution to $\vec{E}_{k}$ from the closest grid intersection and its nearest neighbors: these terms are calculated exactly using Eq.\,\eqref{eq.electro}. Recently, Barros \textit{et al.} proposed an alternative method to numerically calculate the electrostatic interaction between dielectric objects\cite{Barros14}.

\underline{Initial configuration:} We obtained the initial configurations of the simulation by dropping the grains freely onto the bottom while switching off the splash function and the external electric field. Grains are all neutral during initialization and there is neither charge exchange nor polarization following collisions. We wait 10$^{6}$ time steps, by which time grain velocities become negligibly small (less than $|\vec{V}^{s}| \times 10^{-3}$, where $\vec{V}^{s}$ is the splash function defined earlier). At this point, we reinstate the full collision, including the splash function, external field, Coulomb interactions, etc. From this point on, it takes about another 10$^{6}$ time steps for the largest system with 1500 grains to reach an asymptotic steady state, meaning a state whose mean charge per grain reaches an asymptote. 

Finally, some grains at the top of the bed acquire sufficient charge to levitate against gravity. As in prior work Ref. \cite{pahtz2010particle}, we remove these grains from the simulation once they lose contact with other grains. To keep the number of grains constant, we re-inject these grains at the bottom of the bed.  We note that injecting charged grains at the bottom of the bed would introduce a spurious electrical current: to prevent this, we neutralize these grains before re-injection. 

To validate the model, we have measured the number of levitated grains and compared to the experimental values reported by P\"ahtz \textit{et al.}\cite{pahtz2010particle}. We obtained the same qualitative dependence on the number of grain layers.

\end{document}